\documentclass{amsart}
\usepackage{amssymb}
\usepackage{amsfonts}
\usepackage{amsmath}
\usepackage[legalpaper,bookmarks=true,colorlinks=true,linkcolor=blue,citecolor=blue]{hyperref}
\usepackage{graphicx}%
\usepackage{fancyhdr}
\usepackage{color}
\usepackage[mathlines]{lineno}
\usepackage{lscape}
\usepackage{epsfig}
\usepackage{natbib}
\usepackage{geometry}
\usepackage{tgbonum}
\fontfamily{qcr}\selectfont

\newtheorem{theorem}{Theorem}
\theoremstyle{plain}

\newtheorem{lemma}{Lemma}

\numberwithin{equation}{section}

\begin{document}
\Large
\title{Uniform Rates of Convergence of Some Representations of Extremes: a first approach}

\begin{abstract} Uniform convergence rates are provided for asymptotic representations of sample extremes. These bounds which are universal in the sense that they do not depend on the extreme value index are meant to be extended to arbitrary samples extremes in coming papers.\medskip

\noindent \textbf{R\'{e}sum\'{e}.} Des vitesses de convergence uniforme, universelles en ce sens qu'elles ne dépendent pas de l'index extrémal, sont fournies pour des représentations asymptotiques des extrêmes d'échantillon. Ces bornes sont à étendre dans le cas général des extrêmes dans des articles à venir.  \medskip

\noindent \noindent $^{\dag}$ Dr. Modou Ngom.\\
Work Affiliation : Ministery of High School (SENEGAL)\\
LERSTAD, Gaston Berger University, Saint-Louis, S\'en\'egal.\newline
ngomodoungom@gmail.com\\

\noindent $^{\dag}$ Dr Tchilabalo Abozou Kpanzou
University of Kara, Kara, Togo\\
Affiliated to LERSTAD, Gaston Berger University, Saint-Louis, Senegal\\
Emails: t.kpanzou@univkara.net; kpanzout@gmail.com\\

\noindent Cherif Mamadou Moctar TRAORE\\
LMA, FST, EDSTM, University of  Technical and Technological Sciences of Bamako (USTTB), Mali.\\
LERSTAD, Gaston Berger University (UGB), Saint-Louis, Senegal.\\
Email : cheriftraore75@yahoo.com, traore.cherif-mamadou-moctar@ugb.edu.sn.\\

\noindent $^{\dag}$ Dr Moumouni DIALLO\\
Université des Sciences Sociale et de Gestion de Bamako ( USSGB)\\
Faculté des Sciences Économiques et de Gestion (FSEG)\\
Email: moudiallo1@gmail.com.\\

\noindent $^{\dag}$ Gane Samb Lo.\\
LERSTAD, Gaston Berger University, Saint-Louis, S\'en\'egal (main affiliation).\newline
LSTA, Pierre and Marie Curie University, Paris VI, France.\newline
AUST - African University of Sciences and Technology, Abuja, Nigeria\\
gane-samb.lo@edu.ugb.sn, gslo@aust.edu.ng, ganesamblo@ganesamblo.net\\
Permanent address : 1178 Evanston Dr NW T3P 0J9,Calgary, Alberta, Canada.\\

\end{abstract}
\maketitle

\section{Introduction} \label{sec1}

\noindent This paper presents a view on the rate of convergence of the univariate extremes of samples in a simple form. Rather than trying to handle the general problem for a distribution function in the extreme domain of attraction, we focus here on the simplest representations for which we give the most precise rates. The results are expected to serve as tools in general for the univariate case and later for the multivariate frame.\\

\noindent In a few words, the extreme value theory started with the univariate case, especially with independent data. Given a sequence of independent and identically distributed random variables $(X_n)_{n\geq 0}$ with common cumulative distribution function  (\textit{cdf)} and defined on the same probability space $(\Omega, \mathcal{A},\mathbb{P})$, the max-stability problem consists of finding possible limits in distribution of the sequences of partial maxima

$$
M_n=max(X_1, ...,X_n), \ n\geq 1,
$$

\bigskip \noindent when appropriately centered and normalized. Precisely, we want to find non-random sequences $(a_n>0)_{n\geq 1} \subset \mathbb{R}$ and $(b_n)_{n\geq 1} \subset \mathbb{R}$ such that $(M_n-b_n)/a_{n}$ converges in distribution to a random variable $Z$ as $n\rightarrow +\infty$, denoted as 
$$
\frac{M_n-b_n}{a_n} \rightsquigarrow Z. \ \ (M)
$$

\noindent This problem was solved around the middle of 20-th century with the contributions of many people, from whom we can cite \cite{gne}, \cite{fis}, \cite{fre}, etc. The following result is usually quoted as the \cite{gne} result since he had the chance to close the characterization theorem. If $Z$ is non-degenerate, that is : $Z$ takes at least two different values, Formula (M) can hold only if $Z$ is one of the three types in terms of its \textit{cdf}: \\

\noindent the Fréchet type with parameter $\gamma \in \Gamma_{1}=\{x \in \mathbb{R}, \ x>0\}$ :

$$
H_{\gamma}(x)\equiv \phi_{\gamma}(x)=\exp(-x^{-1/\gamma})1_{(x\geq 0)},
$$   

\bigskip \noindent the Weibull type with parameter $\gamma \in \Gamma_{2}=\{x \in \mathbb{R}, \ x<0\}$ :

$$
H_{\gamma}(x)\equiv \psi_{\gamma}(x)=\exp((-x)^{-1/\gamma})1_{(x<0)} + 1_{(x\geq 0)}
$$   

\bigskip \noindent and the Gumbel type with parameter $\gamma \in \Gamma_{0}=\{0\}$ :

$$
H_{\gamma}(x)\equiv\Lambda(x)=\exp(-e^{-x}), \ \ \ x \in \mathbb{R}.
$$   

\bigskip \noindent By type of distribution, we mean that any non-degenerate $Z$ in Formula (M) have a \textit{cdf} $F_Z$ which is of the form $F_Z(x)=H_{\gamma}(Ax+B)$, where $0<A \in \mathbb{R}$, $B \in \mathbb{R}$, and $\gamma \in \mathbb{R}$.\\

\noindent A modern account of the theory, including statistical estimation, can be found in \cite{gal}, \cite{haa}, \cite{haa06}, \cite{emb}, \cite{res}, \cite{bei}, etc. This theory is part of the general weak convergence which is thoroughly treated in \cite{bil} and \cite{vaart}. But all our needs in extreme value theory and in weak convergence are already gathered in \cite{ips-wcrv-ang} and \cite{ips-uevt-ang}, which are currently cited in the paper.\\

\bigskip \noindent We denote by $Z_1=Fr(\gamma)$, $Z_2=Wei(\gamma)$ and $Z_0=\Lambda$ random variables admitting the above \textit{cdf}'s respectively. We remark that the \textit{cdf}'s $H_{\gamma}$ are continuous, and the \textit{cdf} of  $(M_n-b_n)/a_{n}$ is $F(a_nx+b_n)^n, \ x\in \mathbb{R}$. By Theorem 3 in Chapter 2 and Point (5) in Chapter 4, Section 1 in \cite{ips-wcrv-ang}, (M) is equivalent to

$$
R_n=\sup_{x \in \mathbb{R}} |F(a_{n}x+b_n)^n - H_{\gamma}(x)| \rightarrow 0 \ as \ n\rightarrow +\infty.
$$

\bigskip \noindent The general problem of the rate of convergence here is to find $R_n$ or to have universal bounds for it.\\

\bigskip \noindent As we already stressed, here we propose results to be used in the general case. To introduce this, suppose that we have a sequence of independent and uniformly distributed random variables $(U_n)_{n\geq 0}$ on $(0,1)$ and let us consider for each the $U_{1,n}=min(U_1,...,U_n)$ for each $n\geq 1$ and define

$$
Z_{n}(1,\gamma)=(nU_{1,n})^{-\gamma}, \ \gamma \in \Gamma_{1}; \ \ Z_{n}(2,\gamma)=-(nU_{1,n})^{-\gamma}, \ \gamma \in \Gamma_{2}
$$

\noindent and 

$$
Z_{n}(0,\gamma)=-\log(nU_{1,n}), \ \gamma \in \Gamma_{0}.
$$

\bigskip \noindent It is straightforward to show that for $\gamma \in \Gamma_{i}$, $Z_{n}(i,\gamma)$ weakly converges to $Fr(\gamma)$, to $Wei(\gamma)$ and to $\Lambda$ accordingly to $i=0$, $i=1$ and $i=2$. But the most important is that, based on Section 3.2 in Chapter 1 in \cite{ips-uevt-ang} for example, if (M) holds with $Z$ non-degenerate, we may always get the following asymptotic representation in distribution

$$
\frac{M_n-b_n}{a_n}= Z_n(i,\gamma) + r_n, \ \text{ with } r_n \rightarrow 0 \ as \ n\rightarrow +\infty, \ (MG)
$$

\bigskip \noindent for some $i \in \{0,1,2\}$ with suitable sequences $(a_n>0)_{n\geq 1} \subset \mathbb{R}$ and $(b_n)_{n\geq 1} \subset \mathbb{R}$.\\

\noindent Because of the general formula (MG), we think it is important to have a special study of the rate of convergence of the  $Z_n(i,\gamma)$'s, which paves the way, further, for the handling of the general rate 
$r_n$.\\

\noindent This is what we are going to do exactly in Section \ref{sec2}.

\section{Rate of convergence of $Z_{n}(i,\gamma)$} \label{sec2}

\noindent We are going to see that a universal bound will be guided by the following rate of convergence

$$
\biggr(e\biggr(1-\frac{1}{n}\biggr)^{n-1}-1\biggr)
$$ 

\bigskip \noindent which is estimated as follows.

\begin{lemma} \label{boundEX} For all $n\geq 2$
$$
0\leq \biggr(e\biggr(1-\frac{1}{n}\biggr)^{n-1}-1\biggr) \leq C_0 \biggr(\frac{1}{2n}+\frac{1}{n^2 \log n}\biggr).
$$

\bigskip \noindent  where $C_0=exp(0.6106739)=1.841673$.
\end{lemma}

\bigskip \noindent \textbf{Proof}. It is given in the Appendix.\\

\noindent The main result of the paper is the following.\\

\begin{theorem} \label{gmc_th01} We have for all $n\geq 2$, for $C_1=2+C_0$,
$$
\sup_{i \in \{0,1,2\}} \sup_{\gamma \in \Gamma_{i}} \sup_{x \in \mathbb{R}} |F_{Z_{n}(i,\gamma)}(x)-F_{Z_i}(x)|\leq \frac{C_1}{4n}
+\frac{C_0}{2n^2 \log n}.
$$
\end{theorem}

\bigskip \noindent \textbf{Proof}. We first deal with the $Z_{n}(1,\gamma ),\gamma >0.$ Set $\alpha =1/\gamma$. We have 

\begin{equation*}
\forall (x\in \mathbb{R}),\text{ }F_{n}(x)=F_{Z_{n}(1,\gamma )}(x)=\left\{ 
\begin{array}{ccc}
\left( 1-\frac{1}{nx^{\alpha }}\right) ^{n} & if & nx^{\alpha }>1 \\ 
0 & otherwise & 
\end{array}%
\right. 
\end{equation*}

\bigskip \noindent  and

\begin{equation*}
\forall (x\in \mathbb{R}),\text{ }f_{n}(x)=F_{Z_{n}(1,\gamma )}^{^{\prime
}}(x)=\left\{ 
\begin{array}{ccc}
\alpha x^{-\alpha -1}\left( 1-\frac{1}{nx^{\alpha }}\right) ^{n-1} & if & 
nx^{\alpha }>1 \\ 
0 & otherwise. & 
\end{array}%
\right. 
\end{equation*}

\bigskip \noindent  Since, as $n\rightarrow +\infty $,%
\begin{equation*}
Z_{n}(1,\gamma )\rightsquigarrow Fr(1/\gamma ),
\end{equation*}

\bigskip \noindent  we have, by Scheff\'{e}'s Theorem (see for example Theorem 4, Chapter 3 in \citet{ips-wcrv-ang}) :

\begin{eqnarray*}
\sup_{x\in \mathbb{R}}\left\vert F_{Z_{n}(1,\gamma )}(x)-\phi _{\gamma
}(x)\right\vert  &\leq &\sup_{B\text{ measurable}}\int_{B}\left\vert
f_{n}(x)-f(x)\right\vert dx \\
&=&\frac{1}{2}\int \left\vert f_{n}(x)-f(x)\right\vert dx,
\end{eqnarray*}

\bigskip \noindent  with

\begin{equation*}
\forall (x\in \mathbb{R}),\text{ }f(x)=F_{Fr(1/\gamma )}^{^{\prime }}(x)=\left\{ 
\begin{array}{ccc}
\alpha x^{-\alpha -1}e^{-x^{-\alpha }} & if & x>0 \\ 
0 & otherwise. & 
\end{array}
\right. 
\end{equation*}

\bigskip \noindent  For short, we also write \ $F_{Fr(1/\gamma )}=F$. Now, we have to compute

\begin{equation}
2a_{n}(\alpha )=\int \left\vert f_{n}(x)-f(x)\right\vert dx. \label{sec0101}
\end{equation}

\bigskip \noindent We have
\begin{eqnarray}
\int \left\vert f_{n}(x)-f(x)\right\vert dx &=&\int_{[0,n^{-1/\alpha
}]}\left\vert f_{n}(x)-f(x)\right\vert dx+\int_{]n^{-1/\alpha },+\infty
\lbrack }\left\vert f_{n}(x)-f(x)\right\vert dx \notag \\
&=&F(n^{-1/\alpha })+\int_{]n^{-1/\alpha },+\infty \lbrack }\left\vert
f_{n}(x)-f(x)\right\vert dx. \label{sec0102}
\end{eqnarray}

\bigskip \noindent  We also have
\begin{equation*}
\int_{]n^{1/\alpha },+\infty \lbrack }\left\vert f_{n}(x)-f(x)\right\vert
dx=\int_{]n^{-1/\alpha },+\infty \lbrack }\alpha x^{-\alpha -1}e^{-x^{-\alpha
}}\left\vert 1-e^{x^{-\alpha }}\left( 1-\frac{1}{nx^{\alpha }}\right)
^{n-1}\right\vert dx. \label{sec0103}
\end{equation*}

\bigskip \noindent Now, let us study
\begin{equation*}
g_{n,\alpha }(x)=e^{x^{-\alpha }}\left( 1-\frac{1}{nx^{\alpha }}\right) ^{n-1}
\end{equation*}

\bigskip \noindent on $]n^{-1/\alpha },+\infty \lbrack$. Differentiating $g_{n,\alpha }(x)$ gives
\begin{equation*}
g_{n,\alpha }^{\prime }(x)=\frac{\alpha}{n}e^{x^{-\alpha}}x^{-\alpha-1}\left( 1-\frac{1}{nx^{\alpha
}}\right) ^{n-2} (x^{-\alpha }-1)\text{ on }]n^{-1/\alpha },+\infty \lbrack .
\end{equation*}

\bigskip \noindent  For $n\geq 2,$  $g_{n,\alpha }^{\prime }(x)$ is positive on $]n^{-1/\alpha},\ 1]$ and negative on \ $[1,+\infty \lbrack $\ so that $g_{n,\alpha }(x)$ is increasing \ $]n^{-1/\alpha },1]$\ \ and decreasing on \ $[1,+\infty\lbrack$. Since for $n\geq 2,$ 
\begin{equation*}
g_{n,\alpha }(1)=e(1-1/n)^{n-1}>1
\end{equation*}

\bigskip \noindent  and, as $n\rightarrow +\infty $,

\begin{equation*}
g_{n,\alpha }(1)\rightarrow 1^{+} \text{ (i.e., by from above 1) },
\end{equation*}

\bigskip \noindent  we have a unique number $x_{n,\alpha }\in ]n^{-1/\alpha },1]$ such that $g_{n,\alpha }(x_{n,\alpha })=1$ and 
\begin{equation*}
0\leq g_{n,\alpha }(x)\leq 1\text{ on  }]n^{-1/\alpha },x_{n,\alpha }]
\end{equation*}

\bigskip \noindent  and for all $x\geq n^{-1/\alpha}$,
\begin{equation*}
1\leq g_{n,\alpha }(x)\leq g_{n,\alpha }(1)=e(1-1/n)^{n-1}.
\end{equation*}

\bigskip \noindent By exploiting the sign of $1-g_{n,\alpha}(x)$ on $]n^{-1/\alpha},1]$ and $]1, \ +\infty]$ respectively, we have

\begin{eqnarray}
&&\int_{]n^{-1/\alpha },+\infty \lbrack }\alpha x^{-\alpha -1}e^{-x^{\alpha
}}\left\vert 1-e^{x^{-\alpha }}\left( 1-\frac{1}{nx^{\alpha }}\right)
^{n-1}\right\vert dx \notag \\
&\leq &\int_{n^{1/\alpha }}^{x_{n,\alpha }}\alpha x^{-\alpha -1}e^{-x^{-\alpha
}}\left( 1-g_{n,\alpha }(x)\right) dx+\int_{]x_{n,\alpha },+\infty \lbrack
}\alpha x^{-\alpha -1}e^{-x^{-\alpha }}\left( g_{n,\alpha }(x)-1\right) dx \notag \\
&=&a_{n}(\alpha ,1)+a_{n}(\alpha ,2). \label{sec0104}
\end{eqnarray}

\bigskip \noindent Next, we have 
\begin{eqnarray}
a_{n}(\alpha ,1) &=&\int_{n^{-1/\alpha }}^{x_{n,\alpha
}}f(x)dx-\int_{n^{-1/\alpha }}^{x_{n,\alpha }}\left( \left( 1-\frac{1}{%
nx^{\alpha }}\right) ^{n}\right) ^{\prime }dx \notag \\
&=&F(x_{n,\alpha })-F(n^{-1/\alpha })-\left( 1-\frac{1}{nx_{n,\alpha
}^{\alpha }}\right) ^{n}. \label{sec0105}
\end{eqnarray}

\bigskip \noindent Combining all, this leads to 
 
\begin{eqnarray}
a_{n}(\alpha ,2)&\leq& \left( g_{n,\alpha }(1)-1\right) (1-F(x_{n,\alpha })) \notag\\
&\leq& \biggr((e(1-1/n)^{n-1}-1)\biggr)\equiv \alpha_n(3) \label{sec0106}
\end{eqnarray}

\bigskip \noindent In total, by putting Formulas \ref{sec0101}-\ref{sec0106}, we get

\begin{eqnarray*}
2\alpha_n(\alpha) &\leq& \exp(-x_{n,\alpha}^{-\alpha})- \left( 1-\frac{1}{nx_{n,\alpha}^{\alpha }}\right) ^{n} + \alpha_n(3)\\
&\leq& \exp(-x_{n,\alpha}^{-\alpha})\biggr(1-   \biggr(\exp(x_{n,\alpha}^{-\alpha})\left( 1-\frac{1}{nx_{n,\alpha}^{\alpha }}\right)^{n-1}\biggr) \left( 1-\frac{1}{nx_{n,\alpha}^{\alpha }}\right)\biggr) + \alpha_n(3) \\
&\leq& \exp(-x_{n,\alpha}^{-\alpha})\biggr(1-   g_{n,\alpha}(x_{n,\alpha}) \left( 1-\frac{1}{nx_{n,\alpha}^{\alpha }}\right)\biggr) + \alpha_n(3) \ (L2)\\
&\leq& \exp(-x_{n,\alpha}^{-\alpha}) \frac{1}{nx_{n,\alpha}^{\alpha }} + \alpha_n(3),\\
&\leq& \frac{1}{n} \biggr(x_{n,\alpha}^{-\alpha} \exp(-x_{n,\alpha}^{-\alpha})\biggr) + \alpha_n(3),\\
\end{eqnarray*}

\bigskip \noindent where we identified $g_{n,\alpha}(x_{n,\alpha})$, which is equal to one, in Line 2 in the above group of formulas. Now, the term between the big parentheses is bounded by the supremum of the function $\ell(x)=x^{-\alpha}e^{-x^{-\alpha}}$ on $(0,1)$ whose derivative, $\alpha x^{-\alpha-1}e^{-x^{-\alpha}}(x^{-\alpha}-1)$, is non-negative on $(0,1)$. Hence our bound is $\ell(1)=1/e$. We conclude that for $n\geq 2$

$$
2\alpha_n(\alpha) \leq \frac{1}{n}+\biggr(e(1-1/n)^{n-1}-1\biggr).
$$

\bigskip \noindent By using Lemma \ref{boundEX}, we finally get

$$
\alpha_n(\alpha)\leq \frac{1}{2n}+C_0 \biggr(\frac{1}{4n}+\frac{2}{n^2 \log n}\biggr).
$$

\bigskip \noindent From there, we use Lemma \ref{boundEX} to conclude the proof for $Z_{n}(1,\gamma)$. To finish for the other cases, it is enough to see that we have the following two formulas:

$$
\sup_{x \in \mathbb{R}} |F_{Z_{n}(2,\gamma)}(x)-F_{Z_2}(x)| = \sup_{x \in \mathbb{R}} |F_{Z_{n}(1,-1/\gamma)}(-x^{-1})-F_{Z_1}(-x^{-1})|
$$

\bigskip \noindent and

$$
\sup_{x \in \mathbb{R}} |F_{Z_{n}(0,\gamma)}(x)-F_{Z_0}(x)| = \sup_{x \in \mathbb{R}} |F_{Z_{n}(1,1)}(e^x)-F_{Z_1}(e^x)|,
$$

\bigskip \noindent which puts an end to the proof.\\

\bigskip \noindent\textbf{Remark}. The bounds provided here are to be used in the general theory of a coming paper.

\section{Appendix}
\noindent \textbf{Proof}. Fix $n\geq 2$. We have

\begin{eqnarray*}
(n-1)\log(1-1/n) = -1 + \sum_{k\geq 1}\left( \frac{1}{k}-\frac{1}{k+1}\right)\frac{1}{n^k},
\end{eqnarray*}

\bigskip \noindent  which in turn implies

$$
1\leq g_{n,\ \alpha}(1)=e\left(1-\frac{1}{n}\right)^{n-1}=\exp\biggr(\sum_{k\geq 1} \frac{1}{k(k+1)n^k}\biggr).  \ (F1)
$$

\bigskip \noindent  By the comparison methods between series with monotonic terms, say $a_n=a(n)$ with the appropriate improper Riemann integral of $a(x)$, we have

\begin{eqnarray*}
\sum_{k\geq 1} \frac{1}{k(k+1)n^k} &\leq& \frac{1}{2n}+\int_{2}^{+\infty} \frac{dx}{n^x}\\
&\leq& \frac{1}{2n}+\frac{1}{n^2 \log n}. \ (F2)
\end{eqnarray*}

\bigskip \noindent  By combining (F1) and (F2), we have, for $0\leq \theta \leq 1$,

\begin{eqnarray*}
0\leq \biggr(e\biggr(1-\frac{1}{n}\biggr)^{n-1}-1\biggr) &\leq& \exp\biggr(\frac{1}{2n}+\frac{1}{n^2 \log n}\biggr)-1\\
&\leq& \biggr(\frac{1}{2n}+\frac{1}{n^2 \log n}\biggr) \exp\biggr(\theta\biggr(\frac{1}{2n}+\frac{1}{n^2 \log n}\biggr)\biggr)\\
&\leq & exp(0.6106739) \biggr(\frac{1}{2n}+\frac{1}{n^2 \log n}\biggr).
\end{eqnarray*}

\end{document}